\documentclass[12pt,epsfig]{iopart}
\usepackage{epsfig}
\begin{document}
\title[Relative stability of zincblende and wurtzite structure in CdX(X = S, Se, Te) series]{Relative
 stability of zincblende and wurtzite structure in CdX(X = S, Se, Te) series - 
A NMTO study }
\author{Soumendu Datta$^{1}$, Tanusri Saha-Dasgupta$^{1}$\footnote[1]{email : tanusri@bose.res.in} 
and D. D. Sarma$^{2,3}$ }
\address{$^1$S.N. Bose National Centre for Basic Sciences,
Kolkata 700 098, India}
\address{$^2$ Centre for Advanced Materials,
Indian Association for the Cultivation of Science, Jadavpur,
Kolkata 700 032, India}
\address{$^3$ Solid State and Structural Chemistry Unit, Indian Institute of Science, Bangalore 560 012, India}
\date{\today}





\section{Introduction}

Tetrahedrally coordinated binary semiconductors of $A^NB^{8-N}$ type having total of $8$ valence electrons 
are compounds of great technological importances. They have been widely studied both theoretically and 
experimentally\cite{wah,jcp}. An interesting aspect of these compounds is that they can belong to either
 cubic or in hexagonal symmetry, namely in zincblende(ZB) or in wurtzite(WZ) structure. The difference
 between the two structures is subtle, both being of tetrahedral coordination, and also the associated 
total energy difference is small, of the order of few tens of meV/atom. These particular aspects have
 made the ZB vs WZ structural stability issue a topic of significant interest that have been discussed 
in literature both in terms of model calculations as well as in terms of first principle calculations\cite{
mozzi, portis, van1, van2, lawa, bloch, zunger1, zunger2, pawlow, zunger3}. Phillips and Van Vechten 
\cite{van1} first gave a well defined prescription in terms of ionicity to predict the crystalline 
phase in binary octet semiconductors of $A^NB^{8-N}$ type, with the help of the dielectric theory. While
 their theory nicely described the transition from four-fold to six-fold coordination and possibly also
 to eight-fold coordination, it was not able to distinguish clearly between structures having the same
 coordination. P. Lawaetz\cite{lawa} investigated the stability of WZ structure more closely. Taking 
into account the short-range elastic forces and long-range Coulomb forces, he correlated the ZB vs WZ
 stability with the deviation of the axial ratio $\frac{c}{a}$ from the ideal value, which was further 
related with a structure-independent charge parameter. However, the results were not fully satisfactory
 because of limited knowledge about the long-range Coulomb-effects in partially 
ionic materials. Christensen et al. \cite{pawlow} studied the structural phase stability of 34 elemental
 and compound semiconductors from first principle electronic structure calculations which involved mostly
 four-fold to six-fold coordination transitions. They studied the
 chemical trends by calculating the valence charge densities using linearized muffin-tin orbital (LMTO)
  based localized basis sets. Their calculated ionicities from tight-binding representation of LMTO 
Hamiltonian, gives better chemical trend than Phillips value. Recently, the ZB - WZ polytypism in binary 
octet semiconductors have been studied by the group of Alex Zunger\cite{zunger3} in terms 
of quantum mechanically defined atomic-orbital radii. They showed a linear scaling between the ZB - WZ 
energy difference and renormalized orbital radii and successfully predicted the exact structural trends 
in most of the octet compounds, except some few cases. The general understanding that emerged out of these
 calculations is that it is the competition between the covalency and the ionicity effects that determines
 the relative stability of ZB vs WZ structures, with covalency favoring ZB structure
 and ionicity favoring wurtzite structure. 

Experimentally, the stable crystal phase of CdS is hexagonal wurtzite structure\cite{bulkcds}.
 For CdSe, zincblende is the stable low temperature phase 
and above a critical temperature, it transforms to wurtzite 
structure\cite{bulkcdse}. On the other hand, CdTe always stabilizes in cubic zincblende 
structure\cite{bulkcdte1}. However, there is a few indication of metastable wurtzite growth for CdTe in
 some special situation\cite{bulkcdte2}. In the present work, we revised the issue of relative 
stability of ZB and WZ in the CdX series, with X = S, Se, Te using Nth-order muffin-tin 
orbital (NMTO) technique\cite{nmto} within the framework of density functional theory (DFT). 
In particular, we employed the ``direct generation of Wannier-like orbitals'' feature of NMTO 
technique for attaining the microscopic understanding in this context.

\section{Crystal structure}
\begin{figure*}[h]
\centerline{\hspace*{0.5cm}
\includegraphics[width=11.5cm,keepaspectratio]{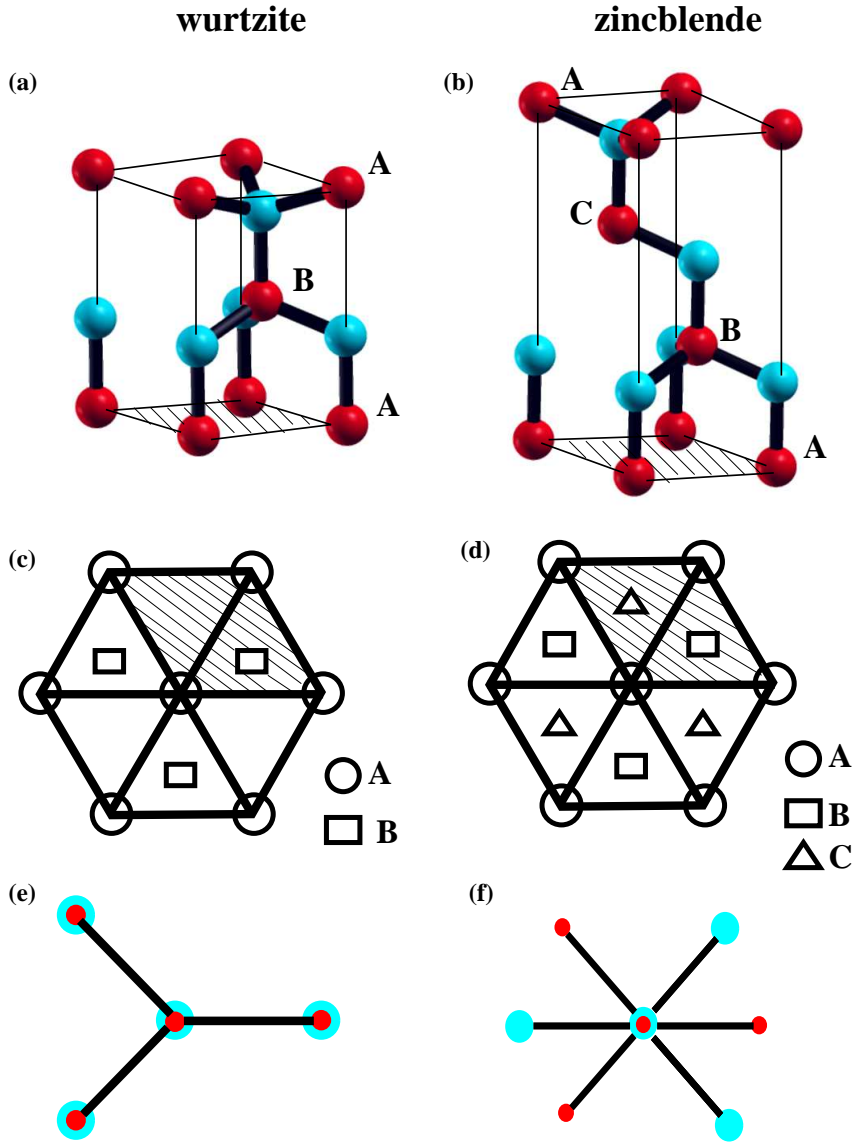}}
\caption{ (a) and (b) show the crystal structure in WZ and ZB symmetries. Two different 
colored atoms denote cation and anion. (c) WZ structure shows the ABAB$\dots$ stacking of atoms, while
 (d) ZB structure shows the ABCABC$\dots$ stacking. {\it Eclipsed} and {\it staggered} conformation of
 atoms along $c$-axis in WZ structure and along [111] body diagonal in ZB structure have been 
shown in (e) and (f) respectively.}
\label{structure}
\end{figure*}

The ZB structure consists of two interpenetrating face centered cubic(FCC) sublattices, one of atom A, 
the other of atom B, displaced from each other along the body diagonal by $\frac{a}{4}$, $a$ being the 
lattice constant for the ZB structure. On the other hand, an ideal WZ structure consists of two
 interpenetrating  hexagonal closed packed(HCP) sublattices, one of atom A, the other of atom B, displaced 
from each other by $\frac{3}{8}c$ along the $c$-axis. These result into two different stacking sequences:
 ABCABC$\dots$ along [111] direction for ZB and ABAB$\dots$ along $c$-axis for WZ. The ZB structure corresponds to the {\it staggered} conformation of atomic arrangement along [111] body diagonal, while the wurtzite 
structure corresponds to the {\it eclipsed} conformation as seen looking down the $c$-axis. The nearest neighbor 
(tetrahedral bond) arrangements in the ZB structure and in the ideal WZ structure are identical. The main difference
 starts to come in the relative position of 3rd nearest neighbors and beyond.  Also the arrangement of the
 distant atoms along the four different tetrahedral bonds are different for a WZ structure. The structural 
differences between these two phases have been shown in Fig. \ref{structure}.

The lower symmetry of the WZ structure allows for a distortion with the $c/a$ ratio deviating from the ideal
 value of $\sqrt\frac{8}{3}$ = 1.633. A further kind of distortion, that is possible in the hexagonal 
structure and forbidden in the cubic, is that the two interpenetrating sublattices are slightly displaced
 from each other so that an atom of one sublattice is no longer at the centroid of the other sublattice.
 In order to capture these distortions, it is necessary to define another internal parameter ``$u$'', which
 is the nearest neighbor distance along $c$-axis, in unit of $c$. The value $u$ in the ideal WZ structure
 is $\frac{3}{8}$ = 0.375. Distortion of $u$ from its ideal value gives rise to a dipole at the center 
of each tetrahedron, containing an atom of opposite kind, the magnitude of which depends upon
 $\Delta u$ [$\Delta u$ = $u$ - 0.375] and the ionicity of the chemical bonds. This effect also lowers 
the ionization energy. The structural parameters of CdS, CdSe and CdTe used in the present calculation are
 listed in Table \ref{table3}.

\begin{table}[h]
\caption{\label{table3}Structural parameters of ZB and WZ phases for CdX (X = S, Se, Te) series. Lattice 
parameters have been taken from Ref. \cite{zunger3} for both ZB and WZ phases of CdS, from Ref. \cite{cdse_zb}
 for ZB CdSe, from Ref. \cite{cdse_wz} for WZ phase of CdSe, from Ref. \cite{bulkcdte1} for ZB CdTe and from 
Ref. \cite{bulkcdte2} for WZ CdTe.}
\begin{indented}
\vskip 0.5cm
\item[]\begin{tabular}{ccccccc}
\br
Compounds &  \multicolumn{5}{c}{WZ} & ZB\\
          & \crule{5}               &       \\ 
          & $a (\AA)$   &$\frac{c}{a}$ & $u$   &    $\Delta(\frac{c}{a})$ & $\Delta u$ & $a (\AA)$  \\
\hline
CdS       & 4.121 &  1.621       & 0.377 &       -0.012   &  0.002 & 5.811 \\
\hline
CdSe      & 4.299 &  1.631       & 0.376 &       -0.002   &  0.001 & 6.077 \\
\hline
CdTe      & 4.57  &  1.637       & 0.375 &        0.004   &  0.000 & 6.492 \\
\br
\end{tabular} 
\end{indented}
\end{table}

It shows that the deviation $\Delta(\frac{c}{a})$ [$\Delta\left(\frac{c}{a}\right)$ = $\frac{c}{a}$ - 1.633] is 
-ve and largest for CdS, while it is +ve for CdTe. The deviation from the ideal $c/a$ ratio is tiny for CdSe. The
deviation $\Delta u$ is found to be opposite in sign to that of  $\Delta(\frac{c}{a})$. A part of the deviation 
$\Delta u$ is expected to come due to nearest neighbor bond bending or bond
 stretching caused by $\Delta(\frac{c}{a})$. The limits on this part, as calculated by Keffer and
 Portis\cite{portis} for AlN, are $u$ = 0.380 to maintain equal nearest neighbour bond lengths and $u$ = 0.372
to maintain equal four tetrahedral angles. However, measured value of $u$ for AlN is 0.385\cite{portis}. This 
means there must be additional contribution in the distortion of $u$ from ideal value, which is coming from
 the long-range Coulomb interaction. NMTO technique based on full self-consistent DFT calculations includes
 both the effect of short-range as
 well as long-range interaction in providing a complete understanding.

\section{Total energy calculations}

In order to check the relative stability of the ZB and WZ structures of the CdS, CdSe and CdTe, we
carried out total energy calculations using the energetically accurate pseudo-potential basis set.  The 
calculations have been performed with ultra-soft pseudopotential\cite{uspp} and local density approximation
 for the exchange-correlation functionals implemented in the Vienna {\it Ab-initio} Simulation 
Package(VASP)\cite{vasp}. The wavefunctions are expanded in the plane wave basis set with a kinetic
 energy cut-off of 250 eV which gives convergence of total energy sufficient to discuss the relative 
stability of various phases. The calculations have been carried out with a k-space grid of 
$11\times11\times\times11$. The obtained total energies and the band gaps are listed in Table 
\ref{table1}. As found, the total energy differences are indeed
 very tiny; of the order of few meV. Our calculations correctly show that CdS to be stable in WZ phase and 
CdTe to be stable in ZB phase. CdSe which is considered as a boarder line case is found to be stable in
 ZB phase in agreement with the published results\cite{zunger3}. The band gaps in WZ and ZB structures, both being direct
 band gaps for CdS, CdSe and CdTe are similar for a given compound. It is found that the band gap decreases
 monotonically in moving from CdS to CdTe in both ZB and WZ symmetries. This monotonic decrease in band gap
indicates the enhanced metallicity across the CdS-CdSe-CdTe series which points to CdTe being more covalent
compared to CdS.

\begin{table}[h]
\caption{\label{table1}Total energy per formula, energy difference $\Delta E$ = $E(ZB)$ - $E(WZ)$ and 
band gaps. + ve sign of $\Delta E$ indicates WZ is more stable than ZB and vice versa. The numbers 
within the parenthesis in the columns of band gap, correspond to 
their experimental values taken from Ref. \cite{bandgap}}
\begin{indented}
\vskip 0.5cm
\item[]\begin{tabular}{cccccc}
\br
Compounds & \multicolumn{2}{c} {Energy(eV/formula)} & $\Delta E$ (meV) & \multicolumn{2}{c}{Band gap(eV)} \\
          &  \crule{2}                              &                  & \crule{2}                      \\       
          &       ZB     &      WZ                  &                  &  ZB        &   WZ                \\
\hline
CdS       &     -6.8650  &  -6.8665                 &    1.5           & 0.94(2.55) &  1.02(2.58)  \\
\hline
CdSe      &   -6.7548    &  -6.7523                 &   -2.5           &  0.77(1.90)&  0.72(1.83)  \\ 
\hline
CdTe      &   -6.0248    &  -6.0221                 &  -2.7            &  0.69(1.60)&  0.65(1.60)  \\ 
\br
\end{tabular}
\end{indented} 
\end{table}

\section{Calculation of ionicity}

As already pointed out, the relative stability between ZB and WZ phases of $A^NB^{8-N}$ semiconductors is 
dictated by the competition between the covalent bonding and the electrostatic energy given by the difference
between the cation and anion energy levels. In a homopolar semiconductors like Si, it is only the first term that
 survives. For a tetrahedrally coordinated semiconductor, it is most natural to think in terms of $sp^3$ hybrids.  Considering the $sp^3$ hybrid energy as 
            $ E_{sp^3} =\frac{E_s + 3E_p}{4}$
 the energy level separation between cation and anion is given by 
$\Delta E_{sp^3}  = E_{sp^3}^c - E_{sp^3}^a$ ;
where $c$ and $a$ denote cation and anion respectively. The hybridization contribution on the other hand is 
related to the hopping integral between the cation and anion $sp^3$ hybrids. It is given by 
 $h$ =  $\frac{1}{4}\left<s^a + p_x^a + p_y^a + p_z^a | H | s^c - p_x^c - p_y^c - p_z^c \right>$ considering the
[111] bond of ZB structure and $h$ =  $\frac{1}{2}\left<s^a + p_z^a | H | s^c - p_z^c \right>$ considering the 
[001] bond of WZ structure where $H$ is the tight binding Hamiltonian in the $sp$ basis of cation and anion.
The ionicity is then defined as

\begin{equation}
 f_i = \frac{(\Delta E_{sp^3})^2}{(\Delta E_{sp^3})^2 + (2h)^2}
\end{equation}

In order to extract the energy level separation and the hopping interaction, we employed the NMTO-downfolding 
technique. In this method, a basis set of localized orbitals is constructed from the exact scattering solutions (partial waves and Hankel functions) for a superposition of short-ranged spherically-symmetric potential wells, 
a so called muffin-tin (MT) approximation to the potential. The basis set is constructed from the scattering solutions at a mesh 
of energies, $\epsilon_0$, $\epsilon_1$, \dots, $\epsilon_N$. At those energies, the set provides the exact
 solutions, while at other energies, $E$, the error is proportional to 
$(E - \epsilon_0)(E - \epsilon_1)\dots(E - \epsilon_N)$. 
The basis set of NMTOs is therefore selective in energy. Moreover, each
NMTO satisfies a specific boundary condition which gives it a specific orbital character and makes it localized.
The NMTO basis set due to its energy selective character is flexible and can be chosen as minimal via the 
downfolding procedure which integrates out the degrees of freedom that are not of interest.The downfolded 
NMTO set spans only the selected set of bands with as few basis functions as there are bands. For the isolated
 set of selected bands the NMTO set spans the Hilbert space of the Waannier functions, that is, the 
orthonormalized NMTOs are the Wannier functions. The Wannier orbitals are therefore generated directly in 
this method, which may be contrasted with the techniques where Wannier functions are generated out of the
 calculated Bloch functions as a post processing step\cite{wannier}. The downfolded Hamiltonian in the Wannier basis
 provides the estimates of onsite energies and the hopping integrals.

       In order to compute the ionicity parameter defined in Eqn. (1), the NMTO calculations have been carried
out with Cd($spd$) and X($sp$) basis. NMTO calculations being not yet implemented in self-consistent form, the
 self consistent calculations have been carried out in the LMTO method. The self
 consistent MT potential out of these LMTO calculations has been used for the constructions of NMTOs, which
 in the present case is the standard LMTO all-electron CdXEE$^\prime$ atomic spheres potential for the ZB 
structure and Cd$_2$X$_2$E$_2$E$_2^\prime$ for the WZ structure. $E$ and $E^\prime$ are two different empty
spheres used to fill the space. The calculated estimates of $\Delta E_{sp^3}$, hopping integral $h$ and 
ionicity parameter $f_i$ for CdX series are listed in Table \ref{table2}. The NMTO approach successfully brings out the 
right trend within the CdX series both for the ZB and the WZ structures, i.e CdS has the maximum ionicity and
 CdTe has the least. For comparison, the results of previous
calculations by Christensen et al.\cite{pawlow} using the tight-binding LMTO approach and that by Phillips
\cite{jcp}
using dielectric theory are also listed. As it is found, while NMTO could bring out the right trend within the 
fine differences, the two previous approaches could not capture it properly.

\begin{table}[h]
\caption{\label{table2}Covalent gap $\Delta$E$_{sp^3}$, hopping term E$_h$ = -2$h$ and ionicity $f_i$ for 
the CdX series in ZB and WZ structure. $f_i$(C) means calculated value of ionicity by Christensen et al.
\cite{pawlow} and $f_i$(P) that of Phillips taken from Ref.\cite{jcp}}
\begin{indented}
\vskip 0.5cm
\item[]\begin{tabular}{ccccccccccc}
\br
Compounds & \multicolumn{2}{c}{$\Delta$E$_{sp^3}$ (eV)} &\multicolumn{2}{c}{$E_h$ = -2$h$ (eV)} &\multicolumn{2}{c}{ $f_i$ }&\multicolumn{2}{c}{ $f_i$(C)} & \multicolumn{2}{c}{$f_i$(P)} \\
          &      \crule{2}  &  \crule{2}     &  \crule{2}     &    \crule{2}    &  \crule{2}   \\
         &   ZB  &  WZ      &   ZB   &  WZ   &  ZB   &   WZ   &   ZB  & WZ      &   ZB   &  WZ   \\
\hline
CdS       & 7.13 & 7.17     &  6.37  & 6.29  & 0.556 &  0.565 & 0.794 & $\dots$ &  0.685 & $\dots$   \\
\hline
CdSe      & 6.70 & 6.69     &  6.16  & 6.04  & 0.542 &  0.551 &  0.841& $\dots$ &  0.699 &  $\dots$  \\
\hline
CdTe      & 5.36 & 5.45     &  5.79  & 5.74  & 0.462 &  0.475 & 0.739 &  $\dots$&  0.717 &  $\dots$ \\
\br
\end{tabular}
\end{indented} 
\end{table}

\section{Microscopic understanding in terms of calculated Wannier functions}

\begin{figure}[h]
\centerline{\hspace*{0.5cm}
\includegraphics[width=13.5cm,keepaspectratio]{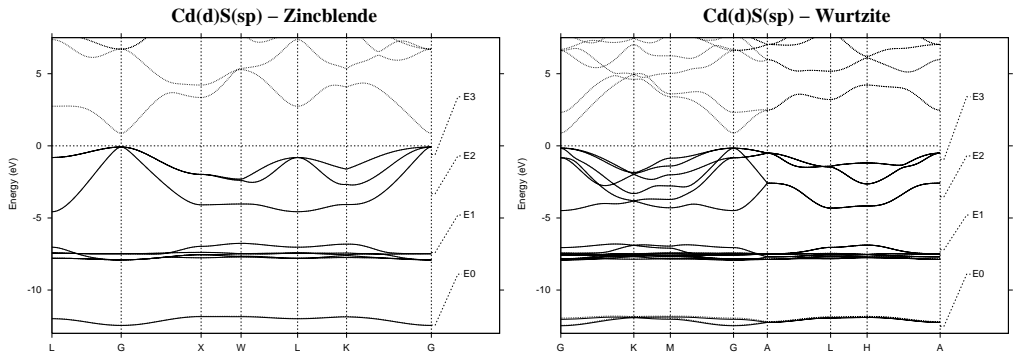}}
\caption{Downfolded bands (thick line) with the basis sets where sp-MTOs are on S atom and
d-MTOs are on Cd atom, are compared with NMTO all bands(thin line) both in ZB(left) and WZ(right)
 structure of CdS. The energy mesh for downfolding have also been shown.}
\label{bands}
\end{figure}

In order to gain further insights, we went a step ahead and constructed the truly minimal NMTO sets with 
the sp-orbitals placed exclusively on the anion(X) site and downfolded all the orbitals at the cation(Cd)
 site, except the cation $d$ orbitals. This gives rise to the basis with only 4 $sp$-orbitals out of
 8 $sp$-orbitals in the ZB unit cell and 8 $sp$-orbitals out of 16 $sp$-orbitals in the WZ unit cell.
 The energy 
points are chosen in the way so as to span only the valence bands. The comparison of the downfolded 
valence-only bands and the full band structure can be made as good as possible by making the energy mesh
 finer and finer. The plot shown in Fig.\ref{bands} for CdS with the choice of 4 energy points already shows the 
downfolded bands to be indistinguishable from the full band structure in the scale of the plot. Similar
 agreements are found also for CdSe and CdTe.

\begin{figure*}[h]
\centerline{\hspace*{0.5cm}
\includegraphics[width=13.5cm,keepaspectratio]{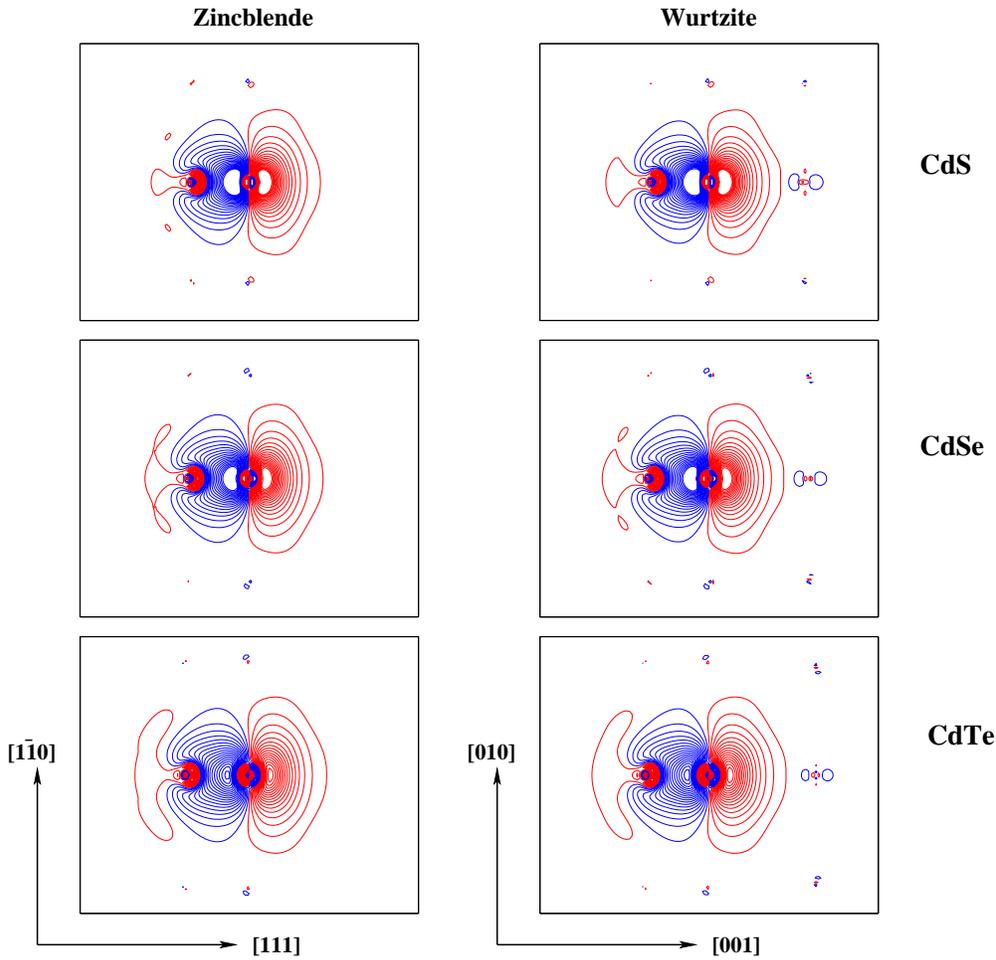}}
\caption{The contour plots show the anion $p[111]$ MTO in ZB structure (left panels)
 and anion $p[001]$ MTO in WZ structure (right panels) of CdX (X = S, Se, Te). The top most panels
 correspond to CdS and the bottom most to CdTe. In each case, 35 contours have been drawn in the 
range -0.15 to 0.15 electrons/Bohr$^3$. From top to bottom, the ionicity decreases and the covalency increases.  }
\label{one_direction}
\end{figure*}

Fig.\ref{one_direction} shows the plot of the orthonormalized p-NMTO (N=3) centered at the anion site and
 pointing to the
neighboring Cd site along the [111] direction for the ZB structure and [001] direction for the WZ structure.
As it is clearly seen, the red lobe at the left-hand side which is a mark of the covalency effect, 
systematically increases in moving from S to Se to Te. 
Had it been plotted for the case of homopolar system like Si which is a perfectly covalent compound, the 
plot would have been perfectly symmetric with the red lobes being symmetric between left hand and the 
right hand side. These plots reconfirm the conclusion that the ionicity decreases and the covalency 
increases across the CdX series. 

\begin{figure}[h]
\centerline{\hspace*{0.5cm}
\includegraphics[width=13.5cm,keepaspectratio]{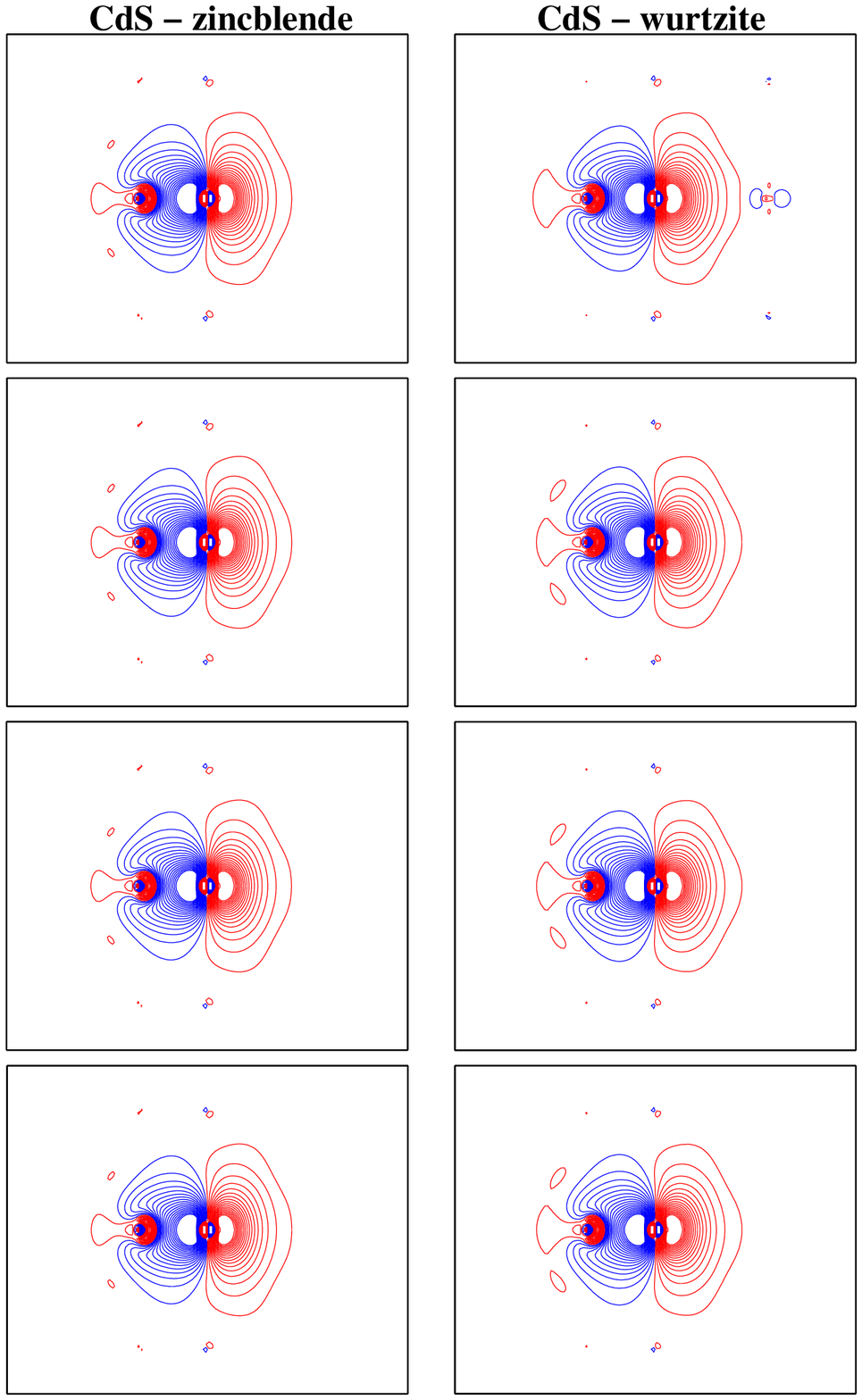}}
\caption{The contour plots show the p-MTO of S along the four nearest neighbor tetrahedral
directions in ZB structure (left panels; from top to bottom they are $p[111]$, $p[1\bar{1}\bar{1}]$,
 $p[\bar{1} 1 \bar{1}]$, $p[\bar{1} \bar{1} 1]$) and in WZ structure (right panels; from top to bottom 
they are p[001], p[1, 0, $\frac{u}{\sqrt{3}}\frac{c}{a}$], p[$\frac{1}{2}$,-$\frac{\sqrt{3}}{2}$, 
 - $\frac{2}{\sqrt{3}} u $$\frac{c}{a}$], p[$\frac{1}{2}$,$\frac{\sqrt{3}}{2}$, 
 - $\frac{2}{\sqrt{3}} u $$\frac{c}{a}$] ) of CdS. The contours chosen are same as in Fig. 
\ref{one_direction}}
\label{cdsall}
\end{figure}

\begin{figure}[h]
\centerline{\hspace*{0.5cm}
\includegraphics[width=15.5cm,keepaspectratio]{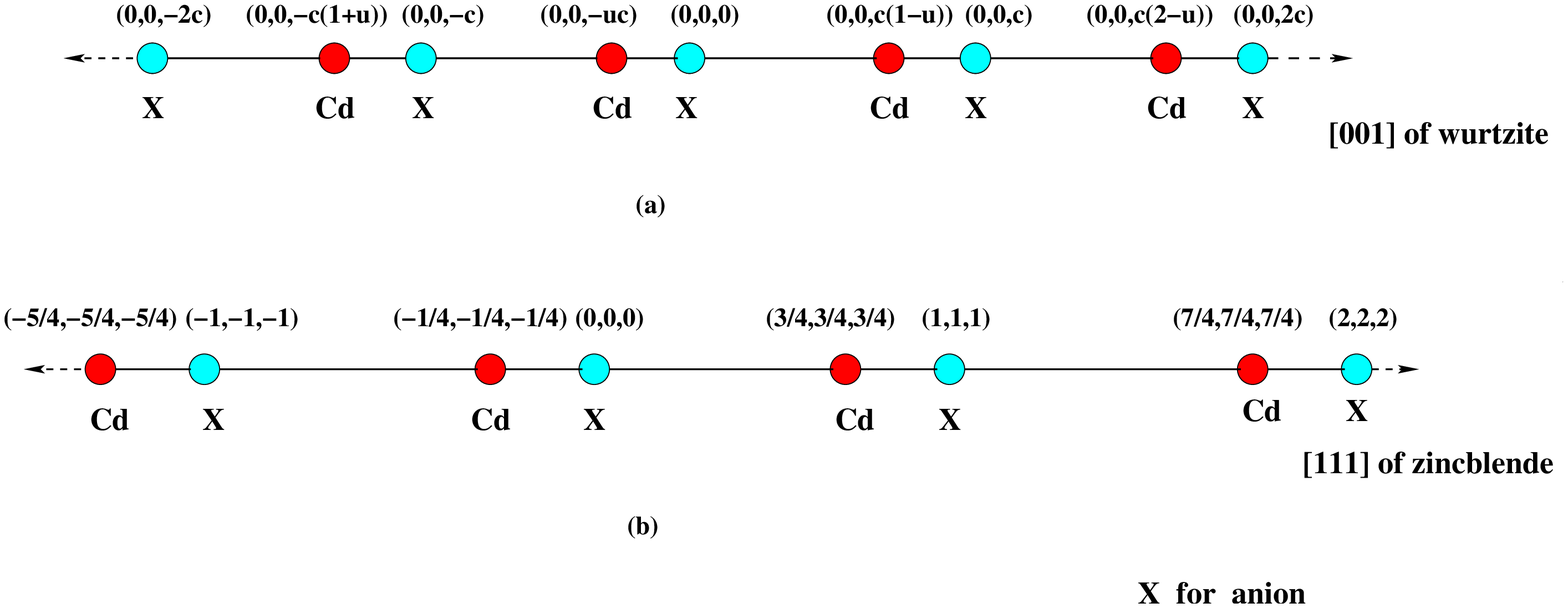}}
\caption{(a) Positions of atoms along [001] direction of wurtzite structure, (b) positions of atoms
 in unit of lattice constant along [111] direction of zincblende structure.}
\label{position}
\end{figure}

  Fig. \ref{cdsall} shows the plots of directed p-NMTO same as in Fig. \ref{one_direction} but for four 
different tetrahedral bond directions of WZ  and ZB structure for CdS. While the four bond-centered p-NMTOs
 look identical for ZB structure, the bond-centered p-NMTO directed along the [001] direction of the WZ 
structure, looks different from the rest of the three. The difference primarily comes from the tail sitting
 at Cd position at $c(1 - u)$ measured from the central anion position at $X(0,0,0)$ along the [001] axis
 ( see Fig. \ref{position}). There is no equivalent neighboring atom along three other directions.

   In order to understand this effect, we have carried out model calculations where $\Delta\left(\frac{c}{a}\right)$ ratios
have been made -0.1, -0.05 and 0.1. The corresponding ``u'' parameter in each case has been obtained by 
total energy minimization fixing the lattice parameters. The deviation in ``u'', $\Delta u$ is found to 
roughly obey the relationship with $\Delta(\frac{c}{a})$ as $\Delta u$ = - 
$\left(\frac{3}{128}\right)^{\frac{1}{2}} \xi \Delta\left(\frac{c}{a}\right)$ where $\xi$, the bond-bending 
parameter, is 2.0, given by P. Lawaetz\cite{lawa}. We computed the ionicity parameter $f_i$ for each of the model systems
 by NMTO downfolding technique. In Table \ref{table4} we list the ionicity parameter for the model WZ systems, together
 with actual CdTe case. As it is evident the -ve deviation from ideal $\left(\frac{c}{a}\right)$ ratio makes the 
compound more ionic as pointed out previously. Changing $\Delta\left(\frac{c}{a}\right)$ ratio from -ve to 
+ve passing through
the ideal $\Delta\left(\frac{c}{a}\right)$ ratio, the ionicity decreases and covalency increases. This is in agreement with 
the structural parameters of CdS, CdSe and CdTe. CdS being the most ionic among the three, shows the largest 
negative deviation in $\left(\frac{c}{a}\right)$ . $\left(\frac{c}{a}\right)$ ratio for CdSe is close to ideal,
while deviation in $\left(\frac{c}{a}\right)$ ratio for CdTe becomes positive.

\begin{figure*}[h]
\centerline{\hspace*{0.5cm}
\includegraphics[width=13.5cm,keepaspectratio]{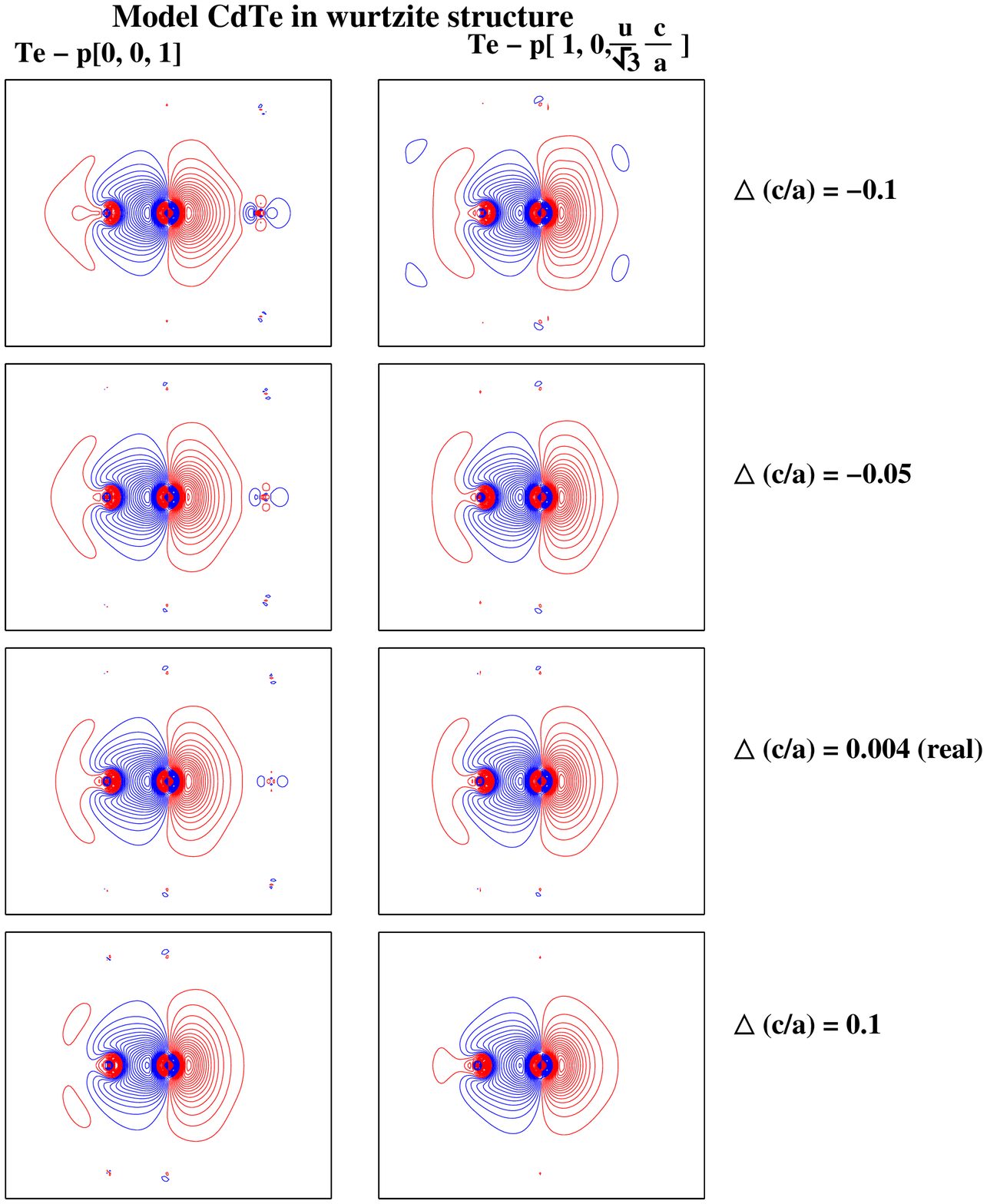}}
\caption{The above contour plots show the p-MTO of Te in WZ structure of model CdTe. The top 
panels correspond to p[001] and bottom panels correspond to p[1, 0, $\frac{u}{\sqrt{3}}\frac{c}{a}$]. From 
left to right ionicity decreases and covalency increases and distribution becomes more isotropic. 
The contours chosen are same as in Fig.\ref{one_direction}}
\label{model}
\end{figure*}

\begin{table}[h]
\caption{\label{table4}Covalent gap $\Delta E_{sp^3}$, hopping term $E_h$ = -2$h$ and ionicity $f_i$ for model
 CdTe in WZ structure, with $\Delta \left(\frac{c}{a}\right)$ and $u$ as given in the first column.}
\begin{indented}
\vskip 0.5cm
\item[]\begin{tabular}{cccc}
\br
CdTe                                                      & $\Delta$E$_{sp^3}$ (eV) &E$_h$ = -2h (eV) &f$_i$  \\
\hline
$\Delta\left(\frac{c}{a}\right)$ = -0.100 and $u$ = 0.4044  & 5.49                     & 5.33            & 0.515  \\
\hline
$\Delta\left(\frac{c}{a}\right)$ = -0.050 and $u$ = 0.3903 & 5.48                     & 5.57            & 0.491  \\
\hline
$\Delta\left(\frac{c}{a}\right)$ = +0.004 and $u$ = 0.375  & 5.45                     & 5.73            & 0.475  \\
\hline
$\Delta\left(\frac{c}{a}\right)$ = +0.100 and $u$ = 0.3432   & 5.45                     & 6.13            & 0.441  \\
\br
\end{tabular} 
\end{indented}
\end{table}

In Fig. \ref{model}, we show the downfolded p-NMTO for the model WZ systems together with actual WZ CdTe
 system along [001]
direction and one of the three other directions, namely [-1, 0, $\frac{u}{\sqrt{3}}$$\frac{c}{a}$ ]. As it is 
evident, changing $\Delta\left(\frac{c}{a}\right)$ from -ve to +ve makes the two p-NMTOs directed along the 
vertical [001] bond and one of the other directions, looking alike, diminishing the tail effect siting at 
Cd[0, 0, c(1 - u)] site. This is driven by the covalency effect which prefers an isotropic arrangement. However,
it cannot be completely achieved within a hexagonal symmetry, a compound with nearly ideal 
$\left(\frac{c}{a}\right)$ or positive $\Delta\left(\frac{c}{a}\right)$ ratio therefore prefers to stabilize 
in ZB symmetry, satisfying the isotropic distribution completely.

\section{Summery}

Using NMTO-downfolding technique we have revisited the problem of ZB vs WZ symmetry in case
 of $A^NB^{8-N}$ semiconductors. In particular, we have considered the CdX series with X =S, Se, Te. Our 
computed ionicity factors using accurate NMTO-downfolding successfully brings out the right trend within the
 CdX series - CdS being most ionic stabilizes in WZ symmetry while CdSe and CdTe being more covalent stabilizes 
in ZB symmetry. Our NMTO constructed Wannier functions corresponding to only valence bands provide nice 
demonstration of this fact. The tendency towards ZB stability is governed by the covalency which prefers isotropic 
nature of the tetrahedral bonds.

\section{Acknowledgment}

S.D thanks CSIR for financial support and TSD acknowledges DST for support through Swarnajayanti fellowship.


\end{document}